\documentclass[12pt]{article}
\usepackage{amsmath,amssymb,graphicx,mathrsfs,hyperref}

\newcommand{\be}{\begin{equation}}
\newcommand{\ee}{\end{equation}}
\newcommand{\bea}{\begin{eqnarray}}
\newcommand{\eea}{\end{eqnarray}}

\def\({\left(} \def\){\right)}
\begin{document}
\title{\vspace{-1.8in}
\vspace{0.3cm} {\ Gravitational entropy and thermodynamics away from the horizon }}
\author{\large Ram Brustein${}^{(1,2)}$,  A.J.M. Medved${}^{(3)}$ \\
 \hspace{-1.5in} \vbox{
 \begin{flushleft}
  $^{\textrm{\normalsize
(1)\ Department of Physics, Ben-Gurion University,
    Beer-Sheva 84105, Israel}}$  $^{\textrm{\normalsize
 (2) CAS, Ludwig-Maximilians-Universit¨at M\"unchen, 80333 M\"unchen, Germany}}$
$^{\textrm{\normalsize (3)  Department of Physics \& Electronics, Rhodes University,
  Grahamstown 6140, South Africa }}$
 \\ \small \hspace{1.7in}
    ramyb@bgu.ac.il,\  j.medved@ru.ac.za
\end{flushleft}
}}
\date{}
\maketitle

\begin{abstract}
We define, by an integral of geometric quantities over a spherical shell of arbitrary radius, an invariant gravitational entropy. This definition relies on defining a gravitational energy and pressure, and it reduces at the horizon of both black branes and black holes to Wald's Noether charge entropy. We support the thermodynamic interpretation of the proposed entropy by showing that, for some cases, the field theory duals of the entropy, energy and pressure are the same as the  corresponding quantities in the field theory. In this context, the  Einstein equations are equivalent to the field theory thermodynamic relation TdS=dE+PdV supplemented by an equation of state.
\end{abstract}

\section{Introduction}

Wald's Noether charge entropy (NCE) \cite{wald1,wald2,jacobson}
is a  conserved geometric charge  that is associated with black hole (BH) horizons. The NCE is related to Bekenstein's area entropy by a specific identification of the dimensionful gravitational coupling \cite{BGH-0712.3206} and  serves as an important conceptual tool for learning about BH  thermodynamics and the holographic paradigm ({\it e.g.}, \cite{ex1,ex2,ex3}). The Noether current is defined on the BH  horizon, whereas the Noether charge is confined to the horizon bifurcation surface and interpreted thermodynamically as the gravitational entropy of the BH. This thermodynamic interpretation relies on relating the horizon Noether charge to the ADM Noether charge, defined at infinity.

In this paper, the gravitational entropy is liberated from the confines of the horizon and its bifurcation surface. To do so, we introduce local, geometric
definitions for an  energy density and a pressure, and  then
apply standard thermodynamics

To be able to define all these quantities, it is necessary to restrict to theories of gravity having a well-defined canonical phase space. We need to consider only UV-complete theories and require that any effective descriptions should reflect their unitarity. (For a detailed discussion on this point, see \cite{prepap}.)
Indeed, we have found that it is only sensible to talk about the gravitational entropy away from the horizon for theories having at most two time derivatives
in the equations of motions.
Any candidate  must then either be a theory of  Lovelock gravity \cite{LL} or, else,  made to mimic this class with a suitable choice of boundary conditions.
The Wald formalism manages to evade such  issues of unitarity because it is valid strictly on
the horizon and  higher-derivative terms can not directly contribute on this Killing surface \cite{BGHM}.

To facilitate the thermodynamic interpretation, we find that it is useful to consider a black brane (BB) with  a Poincar\'e-invariant horizon in an asymptotically  anti-de Sitter (AdS) spacetime.
In this case, a dual field theory is hosted  at the AdS outer boundary \cite{Maldacena1,Witten,Maldacena}.
An  asymptotically flat spacetime and/or a  spherical horizon topology may still be considered but then, in such cases,
a clear thermodynamic interpretation is not always so forthcoming. Much of the supporting  analysis closely follows that of \cite{BwithG}, and so we defer many  calculations and  caveats to this article.

\section{Gravitational entropy away from a horizon}

We wish to propose a  definition for the  gravitational entropy of a region of spacetime. Our concept of a gravitational entropy is that it should be defined in terms of geometric and covariant quantities and should satisfy relations analogous to the first law (and also the second law). Our idea is to define first the energy density $\rho$ and pressure $p$, and then use the equilibrium relation $sT=\rho+p$ to define the product of the entropy density $s$ and the temperature $T$. In symmetric enough cases, such as at a stationary Killing horizon or via further inputs and boundary conditions, it is possible to factor the product and define separately $s$ and $T$. Whereas Wald exploited the former situation, we are
considering the second.

For simplicity, we restrict considerations to $d+1$-dimensional spacetimes that are either asymptotically flat or asymptotically AdS and can be decomposed into $d$-dimensional radial slices with either spherical or planar topologies.
The line element can then be decomposed as
$
\; ds^{2}=N^{2}\,dr^{2}+h_{\alpha\beta}\,dx^{\alpha}\,dx^{\beta}\;,
$
with Roman/Greek indices denoting spacetime/hypersurface
directions. Also, $\;\sqrt{-g}=N\sqrt{-h}\;$.

In these cases, we turn to the Brown--York tensor \cite{BY,stuff} to define the energy density and pressure,
$
T^{ab}\;=\;\lim\limits_{r\to\infty}\frac{2}{\sqrt{-h}}\left(\delta I/\delta {\dot h}_{ab}\right)\;,
$
where  $\;I=\int d^Dx \sqrt{-g}{\cal L}\;$ is the action of the gravitational theory with Lagrangian ${\cal L}$ and a dot denotes a derivative with respect to $r$. As this tensor is generically divergent, a renormalization procedure is required. For an asymptotically flat spacetime, one subtracts the reference Minkowski background, whereas standard procedures of holographic renormalization \cite{skenderis,papa} should be applied when the spacetime is  asymptotically AdS. Thus, the regularized Brown--York tensor has a well-defined  meaning for any spacetime with a timelike or null infinity.

Our next step is to move this tensor (once renormalized) from $r\to\infty$ to a finite radius, $R$. The  resulting tensor
$\widehat{T}^{ab}(R)$ is, like the Brown--York tensor,  precisely defined and finite. As we are dropping terms that diverge at infinity but are otherwise finite, the tensor $\widehat{T}^{ab}(R)$ is a projection of the boundary tensor into the bulk and not a
reapplication of the Brown--York formula at another radius.

Having identified an energy--momentum stress tensor,  we associate  a local energy density and pressure with $\widehat{T}^{ab}(R)$ and identify these as ``gravitational'' densities.  Then, respectively,
\be
\rho(R) \;=\; -\Big.\sqrt{-g}\frac{\delta(r-R)}{\sqrt{g_{rr}}}
\widehat{T}^t_{\ t}\;,
\ee
\be
p(R) \;=\;\Big.\sqrt{-g}\frac{\delta(r-R)}{\sqrt{g_{rr}}} \widehat{T}^x_{\ x}\;.
\ee
It is implicitly
assumed that the densities are to be  evaluated at fixed time in addition
to fixed radius. Hence, we can also obtain a gravitational energy
$E(R)$
by integrating $\rho(R)$ over the spatial coordinates, which amounts to integrating over a spherical shell.  A straightforward calculation shows that, for a Schwarzschild geometry,  $E(R)$
simply gives back the ADM mass.

The standard thermodynamic relation $Ts=\rho+p$
then implies that a gravitational entropy density $s$ and its associated temperature $T$
can also  be defined:
\be
\left.Ts\right|_{r=R}\;=\; -\Big.\sqrt{-g}
\left[ \widehat{T}^t_{\ t}
-\widehat{T}^x_{\ x}\right] \frac{\delta(r-R)}{\sqrt{g_{rr}}}\;.
\label{entxx}
\ee
We can, like above, obtain a gravitational  entropy $S(R)$ by integrating $s(R)$ over a spherical shell. We interpret $s$ as the gravitational entropy of the region $r\le R$.

A prerequisite for identifying the entropy is a clean separation between $s$ and $T$ at any given radius. This is not generically possible but rather requires a high degree of symmetry.
Thus, it is more meaningful to consider the composite
quantity $\Theta\equiv Ts\;$, as we will often do.
In the case of BH and BB geometries, we can identify $T$
with the Hawking temperature. Given that the gravity theory is Einstein's (or any other two-derivative theory
that obeys the equivalence principle), the above definition for
$S(R)$ then  gives back the Bekenstein--Hawking entropy as  $R$ tends to the horizon. The validity of this outcome  will be  made clear by the analysis to follow.

And so a thermodynamic interpretation emerges from the mechanical description. From this point of view, the Einstein equations inform the mechanical interpreter how to change the bulk geometry such that the thermodynamic framework  remains valid at arbitrary radial positions.

\section{Gravitational entropy and thermodynamics away from the horizon for black branes in AdS}

\subsection{Background and conventions}

We begin here with  a $d+1$-dimensional stationary BB  in  an (asymptotically) AdS spacetime
and  the following ansatz for the metric:
$
\;ds^{2}= g_{tt}\,dt^{2}+g_{rr}\,dr^{2}+g_{xx}\,dx^{i}\,dx_{i}\;$, $\;i=1,\dots,d-1\;,
$
with the horizon of the BB and the AdS outer boundary
located at $\;r=r_h\;$ and $\;r\to\infty\;$ respectively. We  sometimes  use ``conformal" metrics for which $g_{rr}=-g_{tt}^{-1}$. All the metric components are assumed to depend only on $r$; meaning that the $\;(t,x_i)\;$ subspace is Poincar\'e invariant.

At the horizon, it is assumed that $g_{tt}$ has a
first-order zero, $g_{rr}$,  a first-order pole and
the others  are finite. It is also assumed that the gravitational  theory asymptotically limits to Einstein gravity at large distance scales.  Since the spacetime
already  asymptotes to pure  AdS
as $\;r\to\infty\;$, the metric has, with the AdS scale set to unity, the asymptotic form
$\;
-g_{tt}\;,\; g^{rr} \to r^2\left[1-\left(\frac{r_h}{r}\right)^d\right]\;$
and $\;g_{xx} \to r^2\;.
$
When we use the  $d+1$-decomposed form for the metric,
then $\;h_{\alpha\beta}=h_{\alpha\beta}(r)\;$.
An index of $x$ or $y$ is used as a representative
of the spatial coordinates $x_i$ and, so, should not be summed over.

When $\;{\cal L}={\cal L}(g_{ab},{\cal R}_{abcd},\psi)\;$
(with ${\cal R}_{abcd}$ denoting the Riemann tensor and
$\psi$, additional matter fields),
Wald's NCE is often expressed as \cite{wald1,wald2,jacobson}
$\;
S_{W}=-\frac{2 \pi}{\kappa}\oint\limits_{\Sigma}
\left(\frac{\partial{\cal L}}{\partial {\cal R}_{abcd}}\right)^{\!\!(0)}
\epsilon_{ab}\epsilon_{cd} \;,
$
where the integration is over a cross-section of the horizon,
the superscript $(0)$ indicates that the evaluation is on-shell and
$\epsilon_{ab}$ is  the horizon bi-normal vector. The surface gravity $\kappa$ is related in the standard way to the Hawking temperature  $\;T_H=\kappa/2\pi\;$, which in terms of the brane metric is given by $\; T_H = -\frac{1}{4\,\pi}
\frac{g_{tt,r}}{\sqrt{-g_{tt}g_{rr}}}|_{r=r_h}\;.$
Because the discussion  is on a BB for
which the ``cross-section'' is infinite, it is more practical
to discuss the entropy density $s_W\;$, as defined
by the integrand of $S_W$.

\subsection{Gravitational entropy away from the horizon}

The Brown--York stress tensor for a Lovelock theory is given by (see Eq.~(\ref{energy-momentum})),
\begin{equation}
T^{tt}\; =\;
4(d-1){\cal X}^{txtx}K_{xx}\;,
\end{equation}
\begin{equation}
T^{xx}\; =\;
4\left[{\cal X}^{xtxt}K_{tt}+(d-2){\cal X}^{xyxy}K_{yy}\right]
\;,
\end{equation}
where we have defined the tensor
$\;
{\cal X}^{abcd}\equiv \frac{\partial{\cal L}}{\partial {\cal R}_{abcd}}\;
$
and  $K_{\alpha\beta}$ denotes the extrinsic
curvature,
\begin{eqnarray}
K^{t}_{\phantom{t}t}&=&\frac{1}{2}\frac{g_{tt,r}}{g_{tt}} \frac{1}{\sqrt{g_{rr}}}\;,
\label{things}\\
K^{x}_{\phantom{x}x}&=&\frac{1}{2}\frac{g_{xx,r}}{g_{xx}} \frac{1}{\sqrt{g_{rr}}}\;.
\label{stuff}
\end{eqnarray}
In terms of the $d+1$-decomposed brane metric,  $\;K_{\alpha\beta} = \dot{h}_{\alpha\beta}/2N\;$.

Now suppose that the ``couplings'' or ${\cal X}$'s do not depend
on the polarization; that is,  ${\cal X}^{ab}_{\;\;\;\;\;ab}$
(with $a\neq b$ and no summation of the indices) is independent of the choice of $a$ and
$b$, as would be the case for Einstein  and
({\em e.g.}) $f({\cal R})$ gravity.  Then,
\be
\Theta(R)\;=\;4\Big.\sqrt{-g}
\left[ K^t_{\ t} {\cal X}^{rt}_{\;\;\;\;\;rt}
- K^x_{\ x}{\cal X}^{rx}_{\;\;\;\;\;rx}\right] \frac{\delta(r-R)}{\sqrt{g_{rr}}}\;.
\label{ent0}
\ee
Like before, it is implicitly assumed that time is also fixed to some specific value, so the entropy is obtained  by integrating the
density on the $d-1$ transverse space ---  just like is done to obtain the NCE.

The entropy density should, on general grounds,
have a structure that is explicitly Poincar\'e invariant. The second term on the r.h.s of Eq.~(\ref{ent0})
vanishes on the horizon, as evident from Eq.~(\ref{stuff}).
(At the horizon, the ${\cal X}$'s are
always order-unity numbers times $1/16 \pi G$.)
Then, on a cross-section  of the  horizon, $\;\Theta(r_h,t_h)= 4\sqrt{-h} K^t_{\ t} {\cal X}^{rt}_{\;\;\;\;\;rt}\;$
and   $\Theta$ reduces to the Wald NCE density,
\be
T_H s_W\;=\;4\left[\sqrt{-g_{tt}}\sigma K^t_{\ t} {\cal X}^{rt}_{\;\;\;\;\;rt}\right]_{r=r_h} \;,
\label{notrel}
\ee
where $\;\sigma=\left(g_{xx}\right)^{\frac{d-1}{2}}\;$ is the area density.

Things are not so simple for theories with
polarization-dependent couplings like ({\em e.g.}) Gauss--Bonnet gravity. A theory whose couplings depend on the  polarization
can also be viewed as Einstein supplemented by additional vector or higher-form fields. So, it is expected that additional thermodynamic parameters (chemical potentials) would be  needed to fully describe the thermodynamics. However, these cases are not conceptually  different from the Einstein case, as long as they have a well-defined phase space and allow for  the separation of the entropy and the temperature; for example, by supporting a radially dependent but Poincar\'e-invariant solution.

\section{Entropy and thermodynamics of the dual field theory}

\subsection{Canonical variables and phase space}

Before identifying the field-theory dual of $\Theta$, we recall some basics.
According to the standard holographic dictionary, the dual of a bulk operator $\Phi$ is defined by
$\;
\langle    \widetilde{\Phi}(x^i)\rangle    =\lim\limits_{r \rightarrow
\infty}\frac{1}{\sqrt{-h}}\Pi_{{\Phi}}(r,x^i)_{ren}\;,
$
where the conjugate momentum $\Pi_{\Phi}$ is, as usual,
$\;
\Pi_{{\Phi}}(r,x^i)\equiv\frac{\partial\,\sqrt{-g}{\cal
L}}{\partial\,\dot{\Phi}(r,x^i)}\;.
$
The subscript $ren$, above, is meant as a
reminder that appropriate procedures of subtraction and
renormalization are required at the boundary. (See, {\it e.g.}, \cite{skenderis,papa} for further details.)

The identification of the canonically conjugate variables or, equivalently, the  Hamiltonian decomposition of the action requires
a well-defined variational principle and, so,  well-formulated boundary conditions, a Cauchy surface and a set of boundary terms. This restriction eliminates
all higher-derivative gravitational theories except for
those of the Lovelock class \cite{MYERS}.

Given that the  canonically conjugate variables are definable, these can be deduced by inspecting the decomposed form of the action $I$, which is given in Eq.~(34)
of \cite{BwithG} (and see \cite{brown} for the original work):
\be
I\;=\;\int\!\! dr \int\!\! dt \int\!\! d^{d-1}x
\sqrt{-h}
\;\biggl(4\;U^{r\beta\gamma r} \;\dot{K}_{\beta\gamma} +2 N U^{\alpha\beta\gamma\delta} K_{\alpha\gamma} K_{\beta\delta} +\cdots
\biggr)\;,
\label{action}
\ee
where $U^{abcd}$ is an auxiliary field which is equal to
${\cal X}^{abcd}$ once the field equations have been imposed. The second term is the Hamiltonian.

The conclusion of \cite{BwithG} is that
${\cal X}^{rt}_{\;\;\;\;\;rt}$ and
$K^t_{\ t}$ are canonical conjugates and, likewise, for the pair
${\cal X}^{rx}_{\;\;\;\;\;rx}$, $K^x_{\ x}$.

\subsection{The field theory dual of the gravitational entropy}

Recalling the standard relationship between the on-shell action and  canonically conjugate variables, we observe that the dual of an operator of the form
$\;\Phi\; \Pi_\Phi\;$ is simply $\;\Pi_\Phi\; \Phi\;$, as the roles of the source and operator get interchanged but the original term in the on-shell action remains the same.

Therefore, the field theory dual to $\Theta$ goes  as
 \be
\widetilde{\Theta}\;=\; \lim_{r\to\infty}
4\Big.\sqrt{-h}\left[ {\cal X}^{rt}_{\;\;\;\;\;rt}K^t_{\ t}
-  {\cal X}^{rx}_{\;\;\;\;\;rx}K^x_{\ x}\right]\Big|_{ren}\;,
\label{ent1}
\ee
subject to the usual renormalization procedures at the boundary.

The gravitational entropy and its field-theory dual both originate from the action term  $\;U^{r\beta\gamma r} \;\dot{K}_{\beta\gamma}\;$ which is of
the usual  adiabatic-invariant form
$J=\;\int^R \Pi_\Phi(r)\ d\Phi/dr\; dr$. Here, the radial coordinate $r$ should be thought of as replacing time, the integration over $t$ and the $x_i$'s is  implied, and we have used that $K_{\alpha\beta}$ commutes with the Hamiltonian
(the second term in the action~(\ref{action})) to replace
the partial derivative  by an exact one. As  neither the Hamiltonian  nor $J$
depend explicitly on $r$, we
can then  follow the standard discussion on adiabatic invariants to convert
$J$ into an integral
on a phase-space surface of  constant energy, $\;J=\oint \Pi_\Phi\ d\Phi\;$.

The adiabatic invariance of $\Theta$ and $\widetilde{\Theta}$ now follows from imposing the on-shell condition, leaving only
a boundary contribution of the form $\;\Pi_\Phi\ \Phi\;$ or $\;\Phi\ \Pi_\Phi$. That the entropy operator and its dual are adiabatic invariants is  not too surprising, as the temperature and entropy already turn up as a canonical pair in most any version of the thermodynamic first law.

\subsection{Thermodynamics at the boundary}

Let us recall  the AdS analogue of the Brown--York
(boundary) stress tensor \cite{BY,stuff}.
Prior to renormalization procedures, this can be
determined from the defining relation
$\;
T^{ab}=\frac{2}{\sqrt{-h}}\frac{\partial
K_{\alpha\beta}}{\partial \dot{h}_{ab}}\frac{\delta I}{\delta
K_{\alpha\beta}}\;$. Since the only dependence on $K_{\alpha\beta}$ comes through the Riemann tensor,
$\;
T^{ab}=\frac{1}{N\sqrt{-h}}
\delta^{a}_{\ \alpha}\delta^{b}_{\ \beta}\frac
{\delta\left(\int\sqrt{-g}{\cal L}\right)}{\delta
{\cal R}_{pquv}}\frac{\partial {\cal R}_{pquv}}
{\partial K_{\alpha\beta}}\;,
$
yielding the final result
$\;
T^{ab}=\delta^{a}_{\ \alpha}\delta^{b}_{\ \beta}\;{\cal X}^{pquv}\;
\frac{\partial {\cal R}_{pquv}}
{\partial K_{\alpha\beta}}$\;.
In a general theory of gravity, calculating $T^{ab}$ would be complicated; however, we have already restricted our attention to Lovelock theories, and for these the task becomes simpler,
\begin{equation}
T^{\alpha\beta}\; =\;
4{\cal X}^{\alpha\gamma\beta\delta}K_{\gamma\delta}\;.
\label{energy-momentum}
\end{equation}

This outcome can be verified by direct calculation but is best understood by way of  the well-known Einstein result,
$
T^{\alpha\beta}_E =
-\frac{1}{8\pi\,G_{d+1}}\left(K^{\alpha\beta} -K h^{\alpha\beta}\right)\;,
$
which  can also be recast as Eq.~(\ref{energy-momentum})
by using  $\;{\cal X}^{abcd}_E=\frac{1}{32\pi G_{d+1}}
\left[g^{ac}g^{bd}-g^{ad}g^{bc}\right]\;$.
The  relative simplicity
of the Einstein stress tensor comes about because terms
in the action with explicit (covariant) derivatives
do not contribute. A term with ({\it e.g.})
$\;X_E^{r\alpha\beta a}\nabla_a K_{\alpha\beta}\;$ can
be  eliminated by integration by parts. Hence, all contributions
to $T^{ab}$ must come from the $\Gamma\Gamma$ part of the Riemann tensor $\;{\cal R}\sim \nabla\Gamma+\Gamma\Gamma\;$. But the same must be true of any other Lovelock theory. In essence, these theories satisfy the defining identity (see Eq.~(3.6) in  \cite{LL})
$\;
\nabla_a {\cal X}^{abcd}\;=\;0\;.
$

Now, to obtain the desired expression for the entropy, we again invoke the standard relations
$\;Ts=\rho +P \;=\;-\sqrt{-h}\left[T^t_{\ t}-T^x_{\ x}\right]\;$.
Denoting this quantity by $\Theta_{FT}$ and then plugging in
the Brown--York tensor~(\ref{energy-momentum}), we obtain
\be
\Theta_{FT}\;=\; \lim_{r\to\infty}
4\sqrt{-h}\Big({\cal X}^{xt}_{\;\;\;\;\;xt}K^t_{\  t}
-  \left[(d-1){\cal X}^{xt}_{\;\;\;\;\;xt}
-(d-2){\cal X}^{xy}_{\;\;\;\;\;xy}\right]K^x_{\ x}\Big)_{ren}\;,
\label{ent2}
\ee
again subject to renormalization procedures.

The operators $\widetilde{\Theta}$ and $\Theta_{FT}$
agree at radial infinity. This is so
because the tensors ${\cal X}^{ab}_{\;\;\;\;\;ab}$ (no summation and
$\;a\neq b\;$) become insensitive to the polarization
as the metric asymptotically approaches
pure AdS space at the boundary.
Adopting the notation
$\;{\cal X}_{\infty}= \lim\limits_{r\to\infty}{\cal X}^{ab}_{\;\;\;\;\;ab}\;$,
we then find that
  \be
\widetilde{\Theta}(r\to\infty)\;=\;\Theta_{FT}(r\to\infty)\;=\;
4\sqrt{-h}\left[ {\cal X}_{\infty} K^t_{\ t}
-  {\cal X}_{\infty}K^x_{\ x}\right]\;.
\label{ent3}
\ee

Eq.~(\ref{ent3}) should be understood as an equivalence between
the two field-theory operators,
$\widetilde{\Theta}$ and
$\Theta_{FT}$. Normally, this would be subject  to the matching of  conformal factors   but, here, the subtraction procedures are   built-in.
This is because, for pure AdS space,
 $\;\rho+p=0\;$ or equivalently $\;K^t_{\ t}
- K^x_{\ x}=0\;$; and so  these operators are, unlike the Brown--York tensor, already finite  at the AdS boundary.

\subsection{Thermodynamics in the bulk revisited}

Comparison between the bulk and the boundary pictures at finite $r$ requires more information. The simplest case is for polarization- and radially independent couplings. From the viewpoint of the boundary theory, this corresponds to a truly conformal field theory whose couplings do not depend on scale. This translates in the bulk to Einstein gravity with its conformal metric, and the thermodynamics can then be described by a temperature, an energy density and a pressure such that the trace of the associated energy-momentum tensor vanishes on any radial slice. That this can be imposed at any radial position is guaranteed by the Einstein equations, which translates to \cite{alexwithL,BwithG}
$\;\frac{d}{dr}\left[\sqrt{-h}\left(K^t_t-K^x_x\right)\right]=0\;$ or
$\;\dot{\Theta}_E=0\;$.

The simplest example of radially dependent  but polarization-independent couplings is an $f({\cal R})$ theory of gravity. This is essentially Einstein gravity coupled
to  a scalar field $\;\psi(r)=f^{\prime}({\cal R})\;$ (a prime denotes a derivative
with respect to the argument).
It is still possible to describe the thermodynamics at different radial positions, but we then need one additional parameter, a chemical potential for the dual of the scalar field.

One finds for an $f({\cal R})$ theory that \cite{BwithG}
\be
\Theta_{f({\cal R})} \;=\; f^{\prime}({\cal R})\Big\{\frac{\sqrt{-h}}{8\pi G}
\left[K^t_t-K^x_x\right]\Big\}\;.
\ee
The term in
the curly brackets is, as already mentioned, a radial invariant and what would normally be  identified
as the Einstein energy density and pressure, $\rho_E+p_E$. We can retain these definitions by adding a new term to the thermodynamic potential
\be
\Theta_{f({\cal R})} \;=\; \rho_E + p_E+ \mu n\;,
\ee
where $n$ is  a ``scalar-charge'' density \cite{sc_ch},
\be
n\;=\; \sqrt{-h}\left[f^{\prime}({\cal R})-1\right]\;
\ee
and the associated  chemical potential $\mu$ takes the form
\be
\mu\;=\;
\frac{1}{8\pi G}\left[K^t_t-K^x_x\right]\;.
\ee
Now, since $\mu n$ is the product of a radial invariant times
a quantity $f^{\prime}({\cal R})-1$ which typically is not, it follows
that
\be
\frac{d\Theta_{f({\cal R})}}{dr}
\;=\;\sqrt{-h} \mu\frac{d\left[{\frac{1}{\sqrt{-h}}}n\right]}{d r}\;,
\ee
which is generally non-vanishing.
Again, the Einstein equations inform the mechanical interpreter how to change the bulk geometry such that the thermodynamic framework  remains valid at arbitrary radial positions. The difference here is that the consistency requires that $\Theta$ changes as a function of $r$.

Finally, it is amusing to consider what happens in the limit of
weak gravity or, equivalently, for $\;r\gg r_h\;$. In this case, we can  use the asymptotic Einstein solution and re-express the metric  in terms of the Newtonian potential $\phi(r)$,
$\;g_{tt}=-r^2\left[1+2\phi(r) +\cdots\right]\;$.
A straightforward calculation then reveals that, to leading order in $\phi$,
$\;
\Theta = 4r^{d+1}{\cal X}_{\infty}\dot{\phi}\;.
$
This is the leading order result for any theory of gravity
that asymptotes to Einstein for $\;r\gg r_h\;$. The association  of $sT$ with the Newtonian  potential is reminiscent of the
``emergent gravity'' proposal \cite{verlinde}. Following
\cite{verlinde}
and identifying an ``Unruh temperature'' $\;T=r^2\dot{\phi}/2\pi\;$,
we then obtain   $\;s=8\pi{\cal X}_{\infty}r^{d-1}=r^{d-1}/4G_{d+1}\;$ or
an off-horizon form of the  Bekenstein--Hawking area law.

\section{Summary}

Let us briefly summarize: Given a UV-complete theory of gravity, we defined a gravitational  energy density and pressure, and
then used these to define  a gravitational entropy.
Next, working in the context of an AdS black brane spacetime,
we identified these geometric constructs with the energy density, pressure and entropy of the field theory at the boundary.

From our results, the following picture emerges: The field theory can be viewed  as a  means for  defining  the thermodynamic quantity $sT$ and the equation of state $\rho(p)$.
The boundary theory also tells us how to connect these via its standard thermodynamic interpretation. Meanwhile,  the bulk geometry endows us with the Einstein equations,
but these are fiducial. They simply instruct a geometrically inclined interpreter on how to change the bulk geometry so as to maintain the validity of  the thermodynamic interpretation at an  arbitrary radial position.

\section*{Acknowledgments}
The research of RB was supported by the Israel Science Foundation
grant no. 239/10. The research of AJMM received support from Rhodes University
and the Claude Leon Merit Award grant no.
CLMA05/2011. AJMM thanks the Arnold Sommerfeld Center for Theoretical Physics at Ludwig-Maximilians-Universitat Muenchen for their hospitality.


\begin{thebibliography}{99}



\bibitem{wald1}
R.~M.~Wald,
  ``Black hole entropy is the Noether charge,''
  Phys.\ Rev.\  D {\bf 48}, 3427 (1993)
  [arXiv:gr-qc/9307038].

\bibitem{wald2}
 V.~Iyer and R.~M.~Wald,
  ``Some properties of Noether charge and a proposal for dynamical black hole
  entropy,''
  Phys.\ Rev.\  D {\bf 50}, 846 (1994)
  [arXiv:gr-qc/9403028].



\bibitem{jacobson}
  T.~Jacobson, G.~Kang and R.~C.~Myers,
  ``On Black Hole Entropy,''
  Phys.\ Rev.\  D {\bf 49}, 6587 (1994)
  [arXiv:gr-qc/9312023].



\bibitem{BGH-0712.3206}
R.~Brustein, D.~Gorbonos and M.~Hadad,
``Wald's entropy is equal to a quarter of the horizon area in units of the
effective gravitational coupling,''
Phys.\ Rev.\ D {\bf 79}, 044025 (2009)
[arXiv:0712.3206 [hep-th]].



\bibitem{ex1}
  A.~Sen,
  ``Black Hole Entropy Function, Attractors and Precision Counting of
  Microstates,''
  Gen.\ Rel.\ Grav.\  {\bf 40}, 2249 (2008)
  [arXiv:0708.1270 [hep-th]].



\bibitem{ex2}
  R.~Brustein and M.~Hadad,
  ``The Einstein equations for generalized theories of gravity and the
  thermodynamic relation $\delta Q = T \delta S$ are equivalent,''
  Phys.\ Rev.\ Lett.\  {\bf 103}, 101301 (2009)
  [Erratum-ibid.\  {\bf 105}, 239902 (2010)]
  [arXiv:0903.0823 [hep-th]].



\bibitem{ex3}
  L.~Y.~Hung, R.~C.~Myers and M.~Smolkin,
  ``On Holographic Entanglement Entropy and Higher Curvature Gravity,''
  JHEP {\bf 1104}, 025 (2011)
  [arXiv:1101.5813 [hep-th]].



\bibitem{prepap}
  R.~Brustein and A.~J.~M.~Medved,
  ``Non-perturbative unitarity constraints on the ratio of shear viscosity to
  entropy density in UV complete theories with a gravity dual,''
Phys. Rev. D {\bf 84}, 126005 (2011);
 [arXiv:1108.5347 [hep-th]].



\bibitem{LL} D. Lovelock, ``The Einstein Tensor And Its Generalizations,''
J. Math. Phys. {\bf 12}, 498, (1971).



\bibitem{BGHM}
  R.~Brustein, D.~Gorbonos, M.~Hadad and A.~J.~M.~Medved,
  ``Evaluating the Wald Entropy from two-derivative terms in quadratic
  actions,''
  Phys.\ Rev.\  D {\bf 84}, 064011 (2011)
  [arXiv:1106.4394 [hep-th]].


\bibitem{Maldacena1}
  J.~M.~Maldacena,
  ``The large N limit of superconformal field theories and supergravity,''
   Adv.\ Theor.\ Math.\ Phys.\  {\bf 2}, 231 (1998)
  [Int.\ J.\ Theor.\ Phys.\  {\bf 38}, 1113 (1999)]
  [arXiv:hep-th/9711200].


\bibitem{Witten}
  E.~Witten,
  ``Anti-de Sitter space and holography,''
  Adv.\ Theor.\ Math.\ Phys.\  {\bf 2}, 253 (1998)
  [arXiv:hep-th/9802150].


\bibitem{Maldacena}
O.~Aharony, S.~S.~Gubser, J.~M.~Maldacena, H.~Ooguri and Y.~Oz,
``Large N field theories, string theory and gravity,''
Phys.\ Rept.\  {\bf 323}, 183 (2000)
[arXiv:hep-th/9905111].







\bibitem{BwithG}
  R.~Brustein and D.~Gorbonos,
  ``The Noether charge entropy in anti-deSitter space and its field theory
  dual,''
  Phys.\ Rev.\  D {\bf 79}, 126003 (2009)
  [arXiv:0902.1553 [hep-th]].


\bibitem{BY}
  J.~D.~Brown and J.~W.~York,
  ``Quasilocal energy and conserved charges derived from the gravitational
  action,''
  Phys.\ Rev.\  D {\bf 47}, 1407 (1993)
  [arXiv:gr-qc/9209012].






\bibitem{stuff} V.~Balasubramanian and P.~Kraus,
``A Stress Tensor for Anti-de Sitter Gravity'',
Commun. Math. Phys. {\bf 208}, 413 (1999)
[arXiv:hep-th/9902121].


\bibitem{skenderis}
  K.~Skenderis,
  ``Lecture notes on holographic renormalization,''
  Class.\ Quant.\ Grav.\  {\bf 19}, 5849 (2002)
  [arXiv:hep-th/0209067].



\bibitem{papa}
  I.~Papadimitriou and K.~Skenderis,
  ``AdS / CFT correspondence and geometry,''
  arXiv:hep-th/0404176.








\bibitem{MYERS}
  R.~C.~Myers,
  ``Higher Derivative Gravity, Surface Terms and String Theory,''
  Phys.\ Rev.\  D {\bf 36}, 392 (1987).






\bibitem{brown}
  J.~D.~Brown,
  ``Black hole entropy and the Hamiltonian formulation of diffeomorphism
  invariant theories,''
  Phys.\ Rev.\  D {\bf 52}, 7011 (1995)
  [arXiv:gr-qc/9506085].

\bibitem{alexwithL}
  A.~Buchel and J.~T.~Liu,
  ``Universality of the shear viscosity in supergravity,''
  Phys.\ Rev.\ Lett.\  {\bf 93}, 090602 (2004)
  [hep-th/0311175].




\bibitem{sc_ch}
  S.~O.~Alexeev and M.~V.~Pomazanov,
  ``Black hole solutions with dilatonic hair in higher curvature gravity,''
  Phys.\ Rev.\ D {\bf 55}, 2110 (1997)
  [hep-th/9605106].






\bibitem{verlinde}
E.~P.~Verlinde,
  ``On the Origin of Gravity and the Laws of Newton,''
  JHEP {\bf 1104}, 029 (2011)
  [arXiv:1001.0785 [hep-th]].



\end{thebibliography}
\end{document}